\documentclass[a4paper,11pt]{article}

\usepackage[labelfont=bf,labelsep=period]{caption}
\usepackage[nodayofweek]{datetime}
\usepackage{enumitem}
\usepackage{float}
\usepackage[margin=2.5cm]{geometry}
\usepackage{graphicx}
\usepackage{hyperref}
\usepackage[utf8]{inputenc}
\usepackage{numbertabbing}
\usepackage{times}
\usepackage{url}
\usepackage{xcolor}
\usepackage{xspace}

\usepackage{array}
\usepackage{subcaption}
\usepackage{tikz}
\usepackage{amsfonts}
\usepackage{booktabs}
\usepackage{makecell}

\usepackage{amsmath} %
\allowdisplaybreaks[2]          %
\usepackage{amssymb} %
\usepackage{amsthm} %
\newtheoremstyle{plain-boldhead}%
  {\topsep}%
  {\topsep}%
  {\itshape}%
  {}%
  {\bfseries}%
  {.}%
  { }%
  {\thmname{#1}\thmnumber{ #2}\thmnote{ (\bfseries #3)}}%
\newtheoremstyle{definition-boldhead}%
  {\topsep}%
  {\topsep}%
  {\normalfont}%
  {}%
  {\bfseries}%
  {.}%
  { }%
  {\thmname{#1}\thmnumber{ #2}\thmnote{ (\bfseries #3)}}%
\theoremstyle{plain-boldhead}

\theoremstyle{definition-boldhead}

\newdateformat{simple}{\THEDAY\ \monthname[\THEMONTH]\ \THEYEAR}
\simple

\floatstyle{ruled}
\newfloat{algo}{htbp}{algo}
\floatname{algo}{Algorithm}

\def \ifempty#1{\def\temp{#1} \ifx\temp\empty }

\newcommand{\etal}{\emph{et al.}}

\newcommand{\CL}{\ensuremath{\mathcal{L}}\xspace}

\newcommand{\CO}{\ensuremath{\mathcal{O}}\xspace}

\providecommand{\rose}[1]{\textcolor{blue}{Rose says: #1}}

\newcommand*\circledfilled[1]{\tikz[baseline=(char.base)]{
            \node[shape=circle, fill=black,text=white,draw,inner sep=1pt,font=\sffamily\footnotesize] (char) {#1};}}
\newcommand*\circled[1]{\tikz[baseline=(char.base)]{
            \node[shape=circle,draw,inner sep=1pt,font=\sffamily\footnotesize] (char) {#1};}}

\graphicspath{ {./images} }

\newcommand{\NAME}{Thetacrypt\xspace} %
\newcommand{\NAMELETTER}{\ensuremath{\Theta}} %
\newcommand{\NAMENETWORK}{\NAMELETTER-network\xspace} %

\begin{document}

\title{\bf \NAME: A Distributed Service for \\Threshold Cryptography}

\author{%
\begin{tabular}{ccc}
  \centering
  \makecell{Mariarosaria Barbaraci\footnotemark[2]\\University of Bern\footnotemark[1]} 
  & \makecell{Noah Schmid\\University of Bern\footnotemark[1]}
  & \makecell{Orestis Alpos\\Common Prefix} \\
  \\
  \multicolumn{3}{c}{
    \begin{tabular}{cc}
      \centering
      \makecell{Michael Senn\\University of Bern\footnotemark[1]} &
      \makecell{Christian Cachin\\University of Bern\footnotemark[1]}
    \end{tabular}
  }
  \\
\end{tabular}
}
 
\footnotetext[1]{Institute of Computer Science, University of Bern,
   Neubr\"{u}ckstrasse 10, 3012 CH-Bern, Switzerland.}
\footnotetext[2]{Contact author. Email: \url{mariarosaria.barbaraci@unibe.ch}}

\date{\today} 

\maketitle

\begin{abstract}\noindent
  Threshold cryptography is a powerful and well-known technique with many applications to systems relying on distributed trust. It has recently emerged also as a solution to challenges in blockchain: frontrunning prevention, managing wallet keys, and generating randomness. This work presents \emph{\NAME}, a versatile library for integrating many threshold schemes into one codebase. It offers a way to easily build distributed systems using threshold cryptography and is agnostic to their implementation language. The architecture of \NAME supports diverse protocols uniformly. The library currently includes six cryptographic schemes that span ciphers, signatures, and randomness generation. The library additionally contains a flexible adapter to an underlying networking layer that provides peer-to-peer communication and a total-order broadcast channel; the latter can be implemented by distributed ledgers, for instance. \NAME serves as a controlled testbed for evaluating the performance of multiple threshold-cryptographic schemes under consistent conditions, showing how the traditional micro benchmarking approach neglects the distributed nature of the protocols and its relevance when considering system performance.
\end{abstract}

\section{Introduction}
\label{sec:intro} 

Distributing cryptographic capabilities within a group of nodes is a well-known technique in resilient systems that goes back to the 1990s~\cite{DBLP:journals/ett/Desmedt94}. Based on \emph{secret sharing}, the field of \emph{threshold cryptography} addresses public-key cryptosystems in which only the collaboration of sufficiently many nodes may produce the cryptographic result. Private keys are protected from leaking to a minority of corrupted nodes, even if they collude, provided most remain intact. Threshold cryptography is an essential and promising technology for securing distributed systems, especially in the recent era of blockchain networks.  
The renewed interest in the technology is confirmed also by the ongoing NIST effort~\cite{nist-tc} to standardize them.

The tension between (static) cryptographic schemes and (dynamic) distributed algorithms makes threshold cryptography inherently complex to build.  We believe that this has hampered its widespread deployment so far.  Implementing cryptographic primitives is difficult in the first place because a minor mistake may compromise the security of an entire solution built on the implementation.  It has therefore become common practice to standardize cryptographic libraries and to use them in a modular way across many applications; OpenSSL is the most prominent example of this development.  Vulnerabilities originating from custom implementations can be reduced, optimizations within a common library, such as eliminating side-channel leakage and using hardware instructions, benefit many applications, and the scrutiny placed on a reusable codebase is amortized across many uses.

Implementing distributed cryptography poses the same technical challenges as implementing ordinary cryptosystems, but with the additional concern of handling the communication among nodes in a distributed network.

Several widely used platforms that integrate threshold cryptosystems with a distributed network have recently emerged, such as the Internet Computer~\cite{DBLP:conf/podc/CamenischDHPS022} developed by DFINITY~\cite{dfinity}, which provides a blockchain-like public computing infrastructure, or DRAND~\cite{drand}, a distributed randomness beacon that produces verifiable and unbiased randomness used by many secure services on the Internet. Moreover, Partisia Blockchain~\cite{partisia23} provides private computation as a service using threshold cryptography. Distributed cryptography is also a prominent method to manage the private keys of cryptocurrency wallets in a resilient way~\cite{dfns, coinbase23}.

However, all these solutions realize threshold cryptosystems \emph{within} their platforms and only for their target applications. Therefore, such cryptographic implementations are specific to the needs of particular deployments and domains. In particular, they cannot easily be decoupled from the underlying communication tools, but a network transport is necessary in distributed cryptography.

To remedy this situation, this work shows how to make threshold cryptography modular and to simplify its deployment.  Our goal is to decouple distributed cryptography from particular applications, such that it can easily be composed with and integrated into diverse distributed protocols. Blockchain platforms offer today significant communication, coordination, and synchronization capabilities; because of their maturity, threshold-cryptographic implementations may rely on them as a black box.  This allows us to focus only on the implementation of the cryptographic schemes.  Several research questions arise in this context:
\begin{itemize}
\item How does one structure a common framework that encompasses multiple types of distributed cryptosystems, such as signatures, encryption, and randomness generation?
\item How can threshold cryptography be integrated with a diverse set of distributed protocols in a modular way?
\item What are the precise assumptions on the communication primitives and which are the interfaces for connecting to existing distributed platforms?
\end{itemize}

To answer these questions, we present \emph{\NAME}, a distributed service for threshold cryptography. It contains implementations of the most prominent distributed cryptographic schemes and organizes them into a hierarchy according to their types and implementations. Moreover, \NAME's modular design allows easy integration of the threshold-cryptographic algorithms with diverse distributed applications. Since \NAME does not depend on a particular networking infrastructure, it may be incorporated with existing platforms that already provide the communication means to execute distributed protocols. For completeness, however, \NAME also comes with its own, minimal communication layer, such that it may be deployed as a standalone service.

\NAME takes care of the challenges related to correctly implementing the cryptographic primitives and handles complex, multi-round distributed schemes modularly.  In this way, threshold cryptography becomes a separate component in distributed applications:  it improves overall security and offers higher flexibility for application developers.  To our knowledge, \NAME is the first system to treat threshold cryptography as a modular component that can be integrated flexibly with multiple applications.

This work describes the structure of \NAME and motivates the need for and applicability of such technology. The codebase presents a layered and modular architecture, favoring easy extensibility and different utilization modes. 
 At a high level, it presents two different APIs that let distributed applications integrate threshold cryptography either as a black box, ignoring protocols' cryptographic details, or in a more granular way, using it as a library of primitives and relying on the assumption of a shared secret. At its core, \NAME operates on a management code that executes tasks through a common interface shared by all implemented protocols.
This interface permits easy and efficient integration of new protocols in the future.  
 The cryptographic core is also modular, functioning as a self-contained library for direct access by specific applications. Language-agnostic interfaces for peer-to-peer and total order broadcast communication have been designed at the network layer to facilitate easy integration with a hosting platform.  
 
 We use our design as a common testbed to fairly compare implemented schemes and evaluate them under different deployment scenarios. 
 We study how the system's performance varies with the number of nodes, network configuration, and schemes' cryptographic characteristics. Our results show that in small deployments, the execution time of a threshold protocol is primarily influenced by the cryptographic assumptions of the scheme. However, as the number of nodes increases, this parameter gets overshadowed by the network latency. 
Further, we introduce novel metrics to quantify performance imbalances among the nodes in threshold-based protocols and analyze the impact on the overall system. These metrics reveal that the trade-off between computationally intensive local processing and network latency contributes to a more balanced system. This perspective complements the traditional micro-benchmarking approach of single cryptographic primitives, providing a more comprehensive view of threshold cryptography performance.

The rest of the paper is structured as follows: Section~\ref{sec:background} reviews threshold cryptography and also highlights practical applications of distributed cryptography in the blockchain space. Section~\ref{sec:thetacrypt} summarizes current approaches with their limitations and then describes the architecture of \NAME; this section also gives the details of the implemented protocols and schemes.
In Section~\ref{sec:evaluation}, we evaluate the performance of the service with varying numbers of nodes and different deployments. Finally, Section~\ref{sec:relwork} explores related work in the literature and Section~\ref{sec:conclusion} concludes the paper.

\section{Background and motivation}
\label{sec:background}

\subsection{Fault-tolerant distributed systems}

Fault tolerance refers to the ability of a system to remain operational even in the presence of faulty components; it is often implemented through replication across a diverse set of \emph{nodes} or \emph{parties} that comprise the system. This work considers the \emph{Byzantine}~\cite{DBLP:journals/toplas/LamportSP82} fault model, where a fraction of the nodes may behave maliciously, and therefore aims at achieving \emph{Byzantine Fault Tolerance~(BFT)}.

BFT protocols are today widely used in blockchain systems. The nodes maintaining a blockchain, called \emph{validators} here, produce the chain together and must agree on the order of transactions and the state of the ledger. To do so, they run a consensus protocol for every new block, implementing total-order broadcast (TOB)~\cite{DBLP:journals/csur/Schneider90}.
The operational model varies according to the assumptions on the network delay and the fraction of Byzantine nodes, which resolves into different blockchain constructions. A wide range of blockchain consensus protocols and implementations exist today~\cite{DBLP:conf/ctrsa/GarayK20}.

\subsection{Threshold cryptography} 

When applied to cryptosystems, fault tolerance means that control over a cryptographic system is distributed among many parties, such that the confidentiality and integrity of the service remains intact despite failure or misbehavior of a subset of the parties. This means that a set of $n$ parties collaborate to jointly compute a cryptographic operation based on a private key. A sufficient number or \emph{threshold} of them, specifically $(t+1)$ or more among the $n$ parties, are needed to carry out an operation successfully. On the other hand, any $t$ or fewer parties must not learn anything about the private key. These schemes leverage the concept of \emph{secret sharing} and constitute the general approach of \emph{threshold cryptography}~\cite{DBLP:journals/ett/Desmedt94}. Threshold cryptography is an active area of research and industrial development, and the Multi-Party Threshold Cryptography project of NIST is currently standardizing threshold-cryptographic protocols~\cite{nist-tc}.

Threshold schemes require a setup phase for the initial distribution of the cryptographic material to the parties. This can either be done by a centralized, trusted dealer or through a distributed key-generation protocol~\cite{DBLP:conf/eurocrypt/Pedersen91a,DBLP:journals/joc/GennaroJKR07}, which is run by the parties themselves. The second approach is more secure but arguably also more complex. A threshold cryptosystem ideally produces the same cryptographic outputs or consumes the same inputs as the corresponding centralized schemes.

The communication requirement is also a crucial factor. \emph{Non-interactive} protocols~\cite{DBLP:journals/joc/ShoupG02,DBLP:journals/joc/BonehLS04,DBLP:conf/globecom/BaekZ03,DBLP:conf/eurocrypt/Shoup00,DBLP:journals/joc/CachinKS05} require one round of communication, allowing asynchronous execution, while \emph{interactive} protocols~\cite{DBLP:conf/acns/CascudoD17,DBLP:conf/sacrypt/KomloG20,DBLP:conf/ccs/RuffingRJSS22,DBLP:conf/acns/GennaroGN16,DBLP:conf/ccs/GennaroG18} require multiple rounds, demanding a certain level of coordination. A typical interface for a non-interactive threshold scheme provides three algorithms: one for generating the share of the partial local cryptographic operation, one to verify partial results, and one to combine the partial results and to assemble the final result.

In \NAME the most relevant kinds of threshold cryptographic schemes are implemented:

\paragraph{Threshold cryptosystem} This is a public-key encryption scheme with a distributed decryption algorithm, where only sufficiently many parties can decrypt a ciphertext. Of particular practical interest are non-interactive schemes that are provably secure against chosen-ciphertext attacks~(CCA)~\cite{DBLP:conf/stoc/NaorY90}.

\paragraph{Threshold signature} This is a public-key digital signature scheme, in which the signing algorithm is distributed. Any group of parties large enough can jointly create a valid signature, while a  strictly smaller group cannot. Desirable properties are \emph{unforgeability}~\cite{DBLP:journals/siamcomp/GoldwasserMR88}, a computationally bounded adversary has a negligible advantage in forging a valid signature, and \emph{robustness}~\cite{DBLP:conf/ccs/CachinKLS02,DBLP:journals/joc/GennaroJKR07}, the protocol will produce the correct output despite misbehavior of faulty parties. 

\paragraph{Distributed randomness} A threshold-random function produces cryptographically secure pseudorandom values based on a shared secret held by the parties. Any group smaller than the threshold cannot distinguish such outputs from truly random values, i.e., its outputs are \emph{pseudorandom}. Many schemes characterize this large category: coin-tossing schemes~\cite{DBLP:journals/joc/CachinKS05}, distributed verifiable random functions~(DVRFs)~\cite{DBLP:conf/eurosp/GalindoLOW21}, and, more generally, distributed randomness beacons~(DRBs)~\cite{DBLP:conf/sp/ChoiMB23,DBLP:conf/sp/DasKI022}. %

\subsection{Motivation}

\paragraph{Practical applications} The widespread adoption of distributed systems and blockchain-based solutions has renewed interest in practical deployments of threshold cryptography. Many current projects integrate threshold schemes with blockchains.

For example, threshold cryptography can be used to \emph{prevent front-running}, where faulty nodes see incoming transactions before these transactions become committed and are executed~\cite{DBLP:conf/sp/DaianGKLZBBJ20}. By encrypting transactions with a service-wide key and letting the nodes decrypt them jointly \emph{after} deciding on their order, this attack can be mitigated~\cite{DBLP:journals/toplas/ReiterB94,DBLP:conf/crypto/CachinKPS01,DBLP:conf/dsn/DuanRZ17}.

Another prominent application today concerns \emph{randomness generation}. Trustworthy, i.e., reliable, non-predictable, and unbiased random values are a key ingredient in many blockchain networks (e.g., for consensus) and applications running on them (e.g., for games). The Internet Computer~\cite{DBLP:conf/podc/CamenischDHPS022}, for instance, uses distributed threshold-cryptographic randomness for its consensus~\cite{DBLP:journals/joc/CachinKS05}.

In high-security contexts, \emph{key management}  often involves multi-factor control, sometimes managed with hardware-security modules (HSMs), but more recently done entirely in software through threshold cryptography.  Moreover, multiple commercial solutions deploy threshold cryptography for managing cryptocurrency wallets today~\cite{dfns, coinbase23}.

\paragraph{Conceptual unification} Despite their practical appeal, threshold-cryptographic schemes come in a wide variety of forms, use diverse communication assumptions, and systems using them often provide an ad-hoc treatment of their features. In particular, the necessary communication among the parties is specialized to the deployment platform, even though the principal requirement (that each party obtains the same sequence of messages) is the same across many deployments.

This level of specialization and integration with specific platforms hampers the wide-spread adoption of threshold cryptography.  If common interfaces for the cryptographic schemes and modular assumptions about the communication platforms were available, a unified view on threshold cryptographic schemes could emerge and accelerate the adoption of this promising technology.

\section{\NAME} 
\label{sec:thetacrypt}

\subsection{Problem statement}
System implementers generally want to rely on robust libraries that provide basic cryptographic operations and, from there, develop more complex protocols. Since writing cryptographic code is complex and requires sophisticated knowledge, multiple client applications can benefit like this from the work put into a common library. 
Compared to the usual algorithms found in cryptographic libraries, threshold cryptography additionally requires the integration and implementation of distributed protocols that handle the communication between the involved parties. However, no common, general framework for threshold cryptography exists so far. This has led to the development of many proprietary implementations in different platforms, which are tailored to and optimized for the specific needs of their target environments. 
Consequently, such solutions often result in cryptographic software modules deeply intertwined with a target system's codebase, making it difficult to port them to other platforms.

The goal of \NAME is to overcome this limitation by providing a general framework that is composable with many distributed platforms. Our solution can be flexibly integrated into different systems without enforcing constraints on a specific network or adversarial model, instead inheriting them from the underlying platform. Moreover, it is highly modular and introduces customizable interfaces to fit multiple deployment configurations.

\subsection{Model}

\NAME operates as a distributed service on a set of \emph{nodes} that are connected together over a network. The network is assumed to provide reliable communication among every pair of nodes, and optionally offer a total-order broadcast primitive among all nodes. No assumptions are made about the network delay.
Each node runs a stateful \emph{\NAME instance} in a dedicated process.  Applications invoke the service at one node through a \emph{remote procedure call (RPC)} in a \emph{client-server} interaction.

\begin{figure}[h!]
  \centering
  \includegraphics[width=0.7\columnwidth]{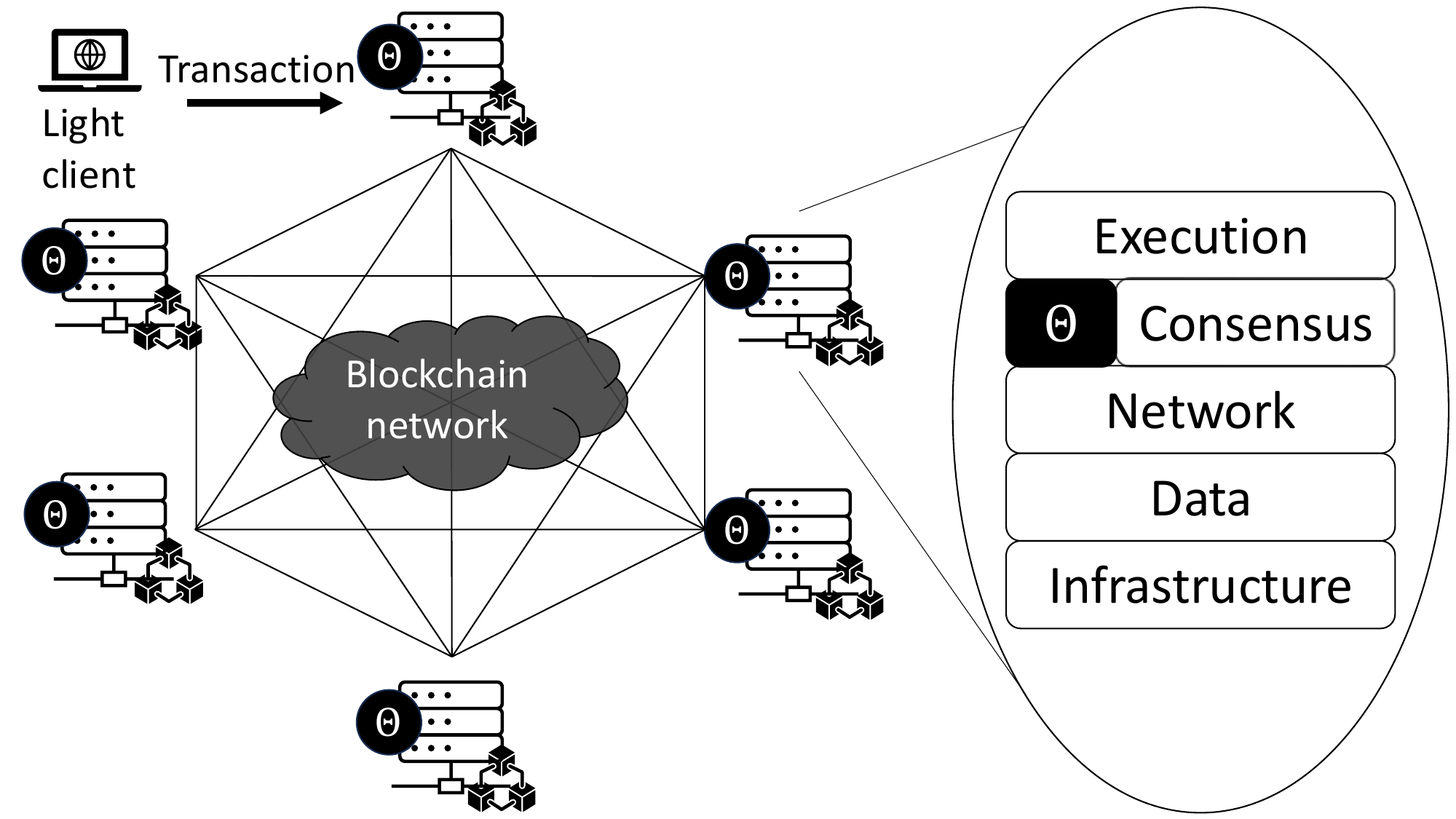}
  \caption{Integration of the \NAME module (\NAMELETTER) in a blockchain
    network, where each network node runs a five-layer blockchain stack.}
  \label{fig:integration}
\end{figure}

The intended deployment of \NAME is in the form of a distributed service coupled with a distributed application that accesses \NAME's operations. The natural setup consists of a fault-tolerant replicated service with as many independent nodes as there are \NAME nodes and with a compatible failure assumption. For instance, one service node may be grouped together with a \NAME node into the same security domain or hosted on the same physical machine.  In this way, $n$ physical nodes can host a service node and a \NAME node each, and up to $t$ of the physical nodes may fail or be corrupted. Every node of the service then issues requests to the \NAME instance locally, and the RPC requests can be authenticated by exploiting the common security context such that only the service node in the same security domain is allowed to issue requests to the \NAME node. 

In this deployment, the distributed application is also responsible for maintaining the service semantics of \NAME: namely, that each instance receives valid inputs, that these inputs respect any particular ordering requirements of the threshold scheme, and that the assumptions of multi-round interactive cryptographic protocols are satisfied.

The envisaged deployment mode of \NAME involves integration with a blockchain network that embodies the state-machine replication paradigm.  In this setting, each node executes an application with deterministic operations, concurrently to the other nodes, and invokes \NAME to obtain service-wide cryptographic results. If integrated with a blockchain platform, applications using \NAME would typically be \emph{smart contracts} that access it through specific API calls from the blockchain validator node. Figure~\ref{fig:integration} illustrates this deployment and shows how the \NAME module (\NAMELETTER) integrates with the blockchain stack.

\subsection{Architecture}

\begin{figure}[h!]
  \centering
    \includegraphics[width=0.8\columnwidth]{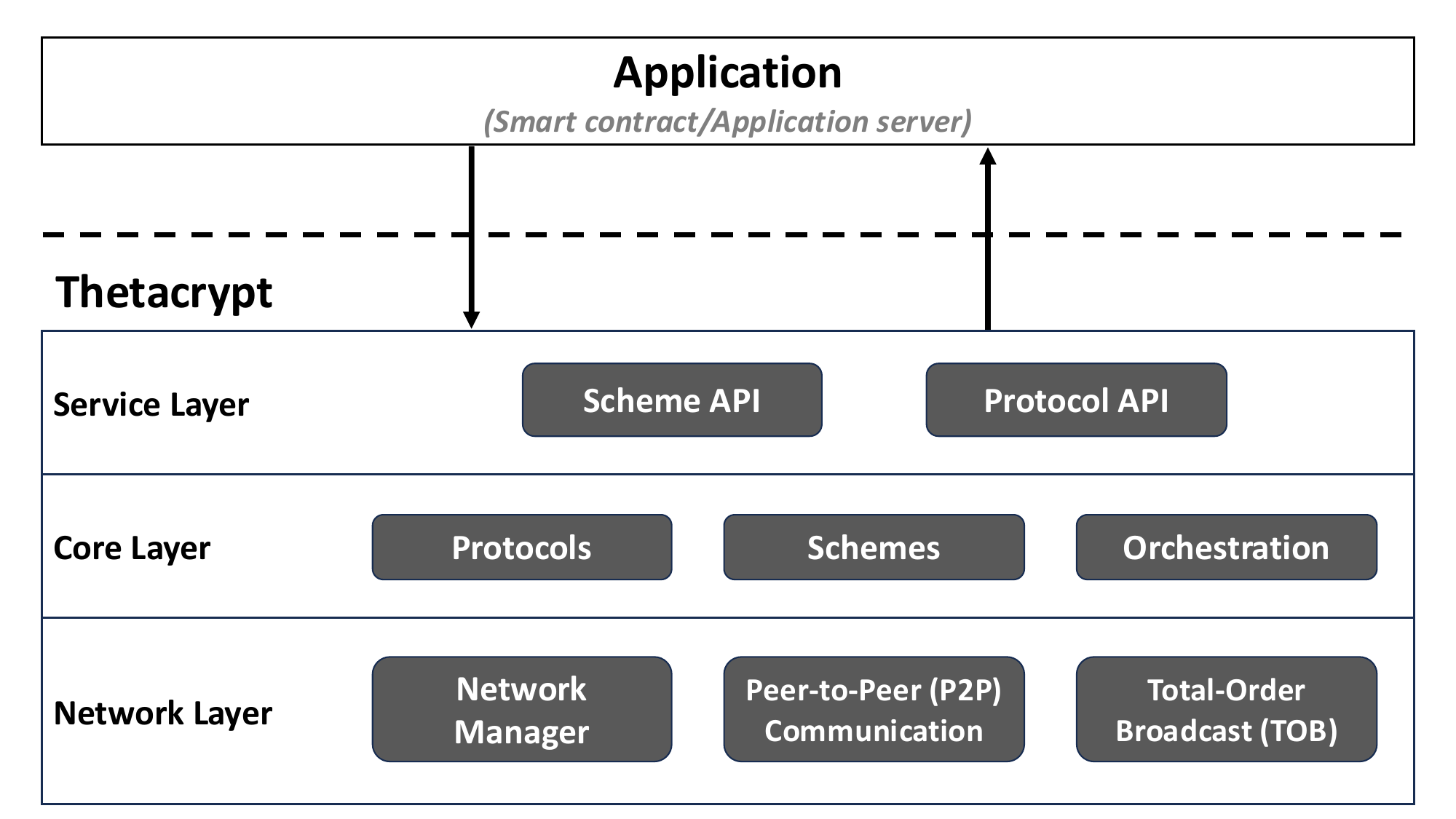}
    \caption{The \NAME architecture consists of three layers. Cryptographic operations are provided by the \emph{schemes module} and they are connected to the network through the \emph{protocols} and \emph{orchestration modules}.}
    \label{fig:stack}
  
\end{figure}

\NAME itself encompasses three layers, as shown in Fig.~\ref{fig:stack}. We first give an overview of all layers and describe them in more detail in the following sections.

Starting from the top, the \emph{service layer} defines the programming interface with two RPC endpoints accessible by the application. The \emph{scheme API} provides direct access to the cryptographic primitives, while the \emph{protocol API} abstracts the execution of threshold protocols.

As the name says, the \emph{core layer} below contains the main logic of \NAME. It is grouped into three modules. The \emph{orchestration module} manages concurrent threads, each performing a threshold cryptographic protocol execution. This module keeps track of the creation, progression, and termination of the protocol instances, communicates with the service layer to return results, and dispatches protocol messages to and from the communication layer. The \emph{protocols module} implements interfaces to access both non-interactive and interactive (multi-round) threshold protocols, abstracting the logic of the protocol away from the particular schemes. Finally, at the heart of \NAME, the \emph{schemes module} provides the cryptographic primitives for the threshold schemes. This module is self-contained and might also be imported as a library directly by other projects.

Finally, the \emph{network layer} handles the communication among the nodes through a dedicated \emph{network manager module}. This module configures the communication and contains the methods to optionally access an underlying network. Specifically, the manager relies on two primary communication methodologies. The \emph{peer-to-peer (P2P) communication module} abstracts internode communication, and the \emph{total-order broadcast (TOB) module} abstracts the integration with an existing replicated service and its total-order broadcast protocol since some cryptographic schemes may depend on it.

\subsection{Service Layer}

The \emph{service layer} exposes the service interface, two RPC endpoints, to an application that integrates threshold cryptographic capabilities into its logic. The \emph{protocol API} encapsulates the translation logic from the public API to the internal protocols implemented in the core layer. 
Additionally, the layer provides the \emph{scheme API} endpoint, allowing the application to access the cryptographic primitives of the \emph{schemes module} directly. This interface is beneficial for applications that require low-level access to cryptographic operations, bypassing the atomic execution of a threshold protocol but requiring more control over when and how the operations are performed.
Currently, the service layer is implemented as a gRPC service~\cite{grpc}. The gRPC framework allows the definition of a service interface using Protocol Buffers~\cite{protobuf}, a serialization toolset that provides a language-agnostic interface. The design's modularity allows the layer to support additional interface formats easily.

\subsection{Core Layer}

The \emph{core layer} represents the main part of \NAME, connecting the cryptographic primitives, the logic behind the protocols, and the management code used to handle key material and execution. At this level, the primary objective is to distinguish  between local computation, characterized by cryptographic operations, and internode communication, required by the distributed nature of threshold schemes. Those two distinct tasks are decoupled: the \emph{protocols module} is responsible for the overall execution logic and the \emph{schemes module} is responsible for the cryptographic details. This design allows for easy integration of additional cryptographic protocols.

For instance, to extend the cryptographic suite with an additional signature scheme, one should work in the \emph{schemes module} providing just the necessary cryptographic primitives, and add the scheme to the list of signature schemes available; the \emph{protocol module} will automatically support the new scheme. On the other hand, to design a new algorithm that takes advantage of the many primitives already present in the suite, one should work exclusively in the \emph{protocol module}, and, if necessary, extend the \emph{schemes module} with the required primitives.

The \emph{orchestration module} is responsible for the management and concurrent execution of multiple protocol instances. It provides the execution environment for the general protocol interface.

\paragraph{Protocols} We identify commonalities in threshold cryptographic protocols and design the \emph{Threshold Round Interface (TRI)} to integrate multiple protocols into a common architecture. The TRI handles the progression of round-based protocols, the finalization of the cryptographic results, operations that compute locally, and computing steps in response to messages received over the network. We refer to a \emph{round} as the local computation performed by one party in response to receiving a message over the network until the party produces a result or a message that may be sent to other parties. This pattern is independent of the number of rounds and suitable for multi-round protocols. 
We present the functions defining the template and how they must be used to correctly execute a generic threshold cryptografic protocol.
  
\begin{itemize}
\item A function \emph{do\_round()} triggers the local computation dictated by the protocol in a particular round. Produces as output a \emph{protocol message} that encapsulates an output of the current round to be forwarded to other parties through the network. Each message indicates whether it is to be transported to other parties using P2P communication or broadcast to all using TOB. This function represents the entry point of the protocol execution. Thus, it is called at the beginning of the protocol (first round) and at the beginning of each new round.

\item Upon receiving a message from the network, every party needs to update the state of the protocol and check the conditions for termination or progression. The function \emph{update(message)}
triggers the update of the current state of the protocol, recording the received message. Based on the new state, the protocol tests the condition for termination or else the condition to move to a new round. If these are not satisfied, it shall wait for further messages. 
 
\item The termination conditions are defined by the function \emph{is\_ready\_to\_finalize()}. If satisfied, the protocol can finalize the computation by executing the \emph{finalize()} function (see below).

\item The condition for progression is provided by the function \emph{is\_ready\_for\_next\_round()}. Typically, this will check if a certain threshold of valid messages has been received from the other parties. If satisfied, the protocol advances to the next round, executing the \emph{do\_round()} function again. 

\item Finally, the function \emph{finalize()} contains the operation to compute the result locally (e.g., assembling partial operations received by a threshold of parties) and give it back to the management code of the orchestration module. 
\end{itemize}

Each specific protocol implementation needs to incorporate such state-machine logic describing protocol steps. It should clearly define the operations that determine a round, the conditions to transition through multiple phases, and how to react to inputs from the network and produce the correct outputs for other parties. Finally, locally compute the final result. The actual cryptographic primitives are provided by the \emph{schemes module}.

\paragraph{Schemes}
\label{sec:schemes}
The \emph{schemes module} performs all cryptographic calculations for the threshold schemes implemented in \NAME. It provides an easy-to-use API and represents the cryptographic core of the service.

\begin{table}[h]
    \centering
      \begin{tabular}{lccc}
      \toprule
      Cryptographic scheme & Reference & Hardness & Verification strategy \\
      \midrule
      Signature & SH00~\cite{DBLP:conf/eurocrypt/Shoup00} & RSA & ZKP \\
      Signature & KG20~\cite{DBLP:conf/sacrypt/KomloG20} & DL & ZKP \\
      Signature & BLS04~\cite{DBLP:journals/joc/BonehLS04} & DL & Pairings \\
      Cipher & SG02~\cite{DBLP:journals/joc/ShoupG02} & DL & ZKP \\
      Cipher & BZ03~\cite{DBLP:conf/globecom/BaekZ03} & DL & Pairings \\
      Randomness & CKS05~\cite{DBLP:journals/joc/CachinKS05} & DL & ZKP \\
    \bottomrule
      \end{tabular}
  \caption{Threshold schemes in \NAME }
  \label{tab:schemes}
  \end{table}
The main approach in the choice of the schemes implemented is related to their importance in the literature, and substantiated security proof. 
Table~\ref{tab:schemes} presents an overview of the implemented schemes. %

We will give a brief description of the different kinds of protocols offered, mentioning which schemes from the literature we implemented and their main characteristics. 

\begin{itemize}
\item \textbf{SG02:} The threshold cryptosystem proposed by Shoup and Gennaro~\cite{DBLP:journals/joc/ShoupG02} is the first non-interactive protocol provably secure against CCA. It relies on the Discrete Logarithm problem. 
Our implementation relies on the Decisional Diffie-Hellman construction (TDH2 in the paper), which is an ElGamal-based~\cite{DBLP:journals/tit/Elgamal85} scheme with zero-knowledge~proof~(ZKP) of language membership to ensure the security of the scheme in the threshold setting.
We apply a hybrid approach~\cite{DBLP:journals/iacr/AbdallaBR99,DBLP:conf/ctrsa/AbdallaBR01} to encrypt a symmetric key under the threshold key and the actual plaintext under the symmetric key. As a symmetric encryption scheme, we use the ChaCha20Poly1305, a stream cipher with a message authentication code.
  
\item \textbf{BZ03:} The scheme by Baek and Zheng~\cite{DBLP:conf/globecom/BaekZ03} shares the desirable properties of SG02, but it is based on groups that satisfy the Gap Diffie-Hellman problem and requires pairing-friendly elliptic curves. Using pairings allows for an efficient ciphertext verification, without the need for ZKP. The implementation uses the same hybrid approach as~SG02.
\item \textbf{SH00:} The first construction of a non-interactive robust threshold signature scheme based on RSA is presented and analyzed by Shoup~\cite{DBLP:conf/eurocrypt/Shoup00}. The signature share generation and the verification algorithm use a ZKP for correctness. We provide an implementation of the scheme based on different sizes of the RSA modulus: namely $512$, $1024$, $2048$, and $4096$.
  
\item \textbf{KG20:} The signature scheme introduced by Komlo and Goldberg~\cite{DBLP:conf/sacrypt/KomloG20}, known as FROST, provides an optimized protocol for threshold Schnorr signatures~\cite{DBLP:journals/joc/Schnorr91}. Due to its randomized nature, the signing algorithm is an interactive protocol: the parties first jointly calculate a random nonce (first round) and then sign (second round). FROST introduces a pre-computation phase that performs the first round in advance for a batch of nonces. This is possible because the nonce is independent of the message to be signed. If precomputations are available, the signing algorithm only needs one round of interaction. In our implementation is possible to run a pre-computation phase to speed up the signing process as well as to run the protocol in a two-rounds fashion.
FROST is the first multi-round protocol to have been implemented in \NAME, and served as a model and test case for the proposed design. Incidentally, FROST is not robust, i.e., actively deviating parties may cause the signature protocol to abort.
  
\item \textbf{BLS04:} The signature scheme proposed by Boneh~\etal~\cite{DBLP:journals/joc/BonehLS04} provides a novel construction that produces short signatures compared to RSA and DSA signatures with approximately the same level of security. 
The scheme uses pairings, thus assuming the Gap Diffie-Hellman problem, which provides, thanks to the bilinear mapping property, an efficient and deterministic verification algorithm. The key homomorphism makes the scheme directly threshold-friendly, both for the generation and verification of signature shares.

\item \textbf{CKS05:} The coin-tossing scheme of Cachin \etal~\cite{DBLP:journals/joc/CachinKS05} produces pseudorandom bit strings. Essentially, the scheme is a distributed function mapping a string to the coin value. The paper proposes two constructions, one from any distributed threshold signature scheme with unique signatures (such as the RSA-based SH00), and the other based on the Diffie-Hellman problem. Both constructions are proved secure in the random-oracle model.  \NAME incorporates the second construction, based on the Diffie-Hellman problem.
It is worth noting that every share of a coin comes with a ZKP for validity, which is a proof of equality of discrete logarithms that ensures the correctness of the share.
\end{itemize}

For non-interactive schemes, we provide a higher-level interface that groups the schemes in the categories of \emph{signature}, \emph{cipher}, and \emph{randomness}. In this way, at the protocol level, and then at the service level, we reduce the complexity of the APIs and parametrize them just with the scheme type and the arithmetic group needed for it. In the implementation, for the elliptic-curve arithmetic, we use MIRACL Core, a multi-lingual and architecturally agnostic cryptographic library~\cite{miracl}.

\begin{figure}[t]
  \centering
    \includegraphics[width=\columnwidth]{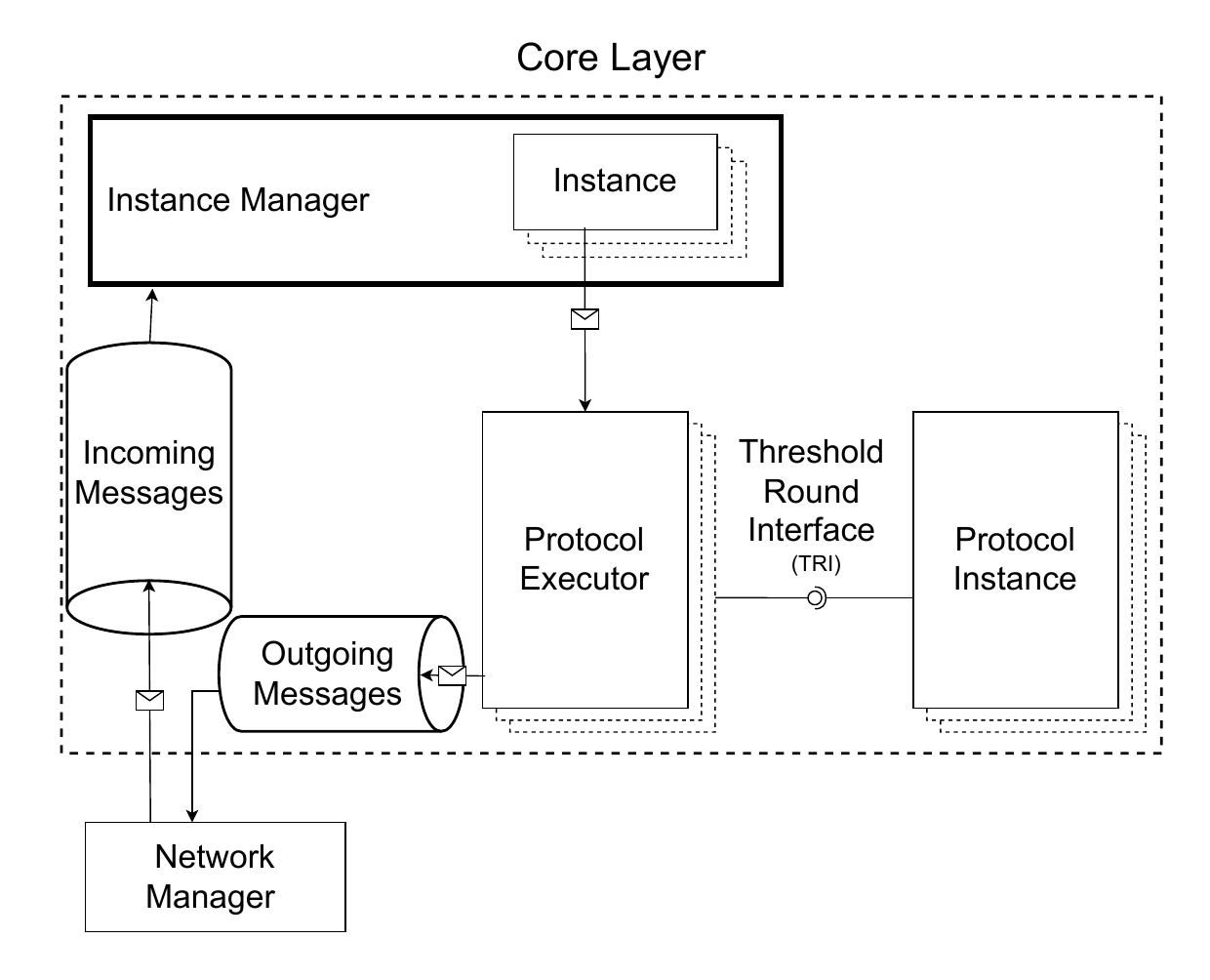}
    \vspace*{-6mm}
    \caption{The \emph{orchestration module} and its components}
    \label{fig:orchestration}
  
\end{figure}

\paragraph{Orchestration} The \emph{orchestration module} implements and exposes the service's internal programming interface. It provides an execution engine that manages multiple protocol instances, tracks related states, and schedules messages to and from the network, as shown in Fig~\ref{fig:orchestration}. Its main component is the \emph{instance manager} that keeps track of the \emph{instances} and is responsible for managing the state of every new instance.

The engine invokes a \emph{protocol executor} for each protocol. This is a state machine that manages the progress of each \emph{protocol instance}. This instance is the actual protocol, implementing the TRI, to which a request refers. The executor is designed to be generic and flexible, allowing the integration of different TRI protocols. It is responsible for ensuring correct execution and proper termination of an instance. Additionally, the executor utilizes a key manager component to access key material.

\subsection{Network Layer}
The \emph{network layer} is designed to serve all the possible deployment configurations of \NAME. A \emph{network manager module} sets up the needed components based on the configuration provided at start-up. By default, the communication layer provides peer-to-peer (P2P) communication, on which, optionally, one can enable end-to-end authentication.
In addition, the network layer can also be configured to access total-order broadcast (TOB). Depending on the scheme, this is a requirement for some multi-round protocols where the parties need to synchronize their communication. 

Hence, the manager works with interfaces: the \emph{P2P interface} and the \emph{TOB interface}. The concrete implementation of network components can either be an actual networking module or a proxy module that delegates the operations to a remote node.

In our implementation, a \emph{P2P component} implements an overlay network running a gossip protocol via libp2p~\cite{libp2p}. Alternatively, we develop two proxy modules, a \emph{P2P proxy module} and a \emph{TOB proxy module}, to realize the integration with an already existing replicated service and delegate the communication to it.

The proxies can request remote services via the communication layer of the underlying system. Thus, our design defines a specific API required by both the proxy and the target systems. While this approach may complicate integration, it enables our service to become part of the host's software stack~(Fig.~\ref{fig:integration}). It simplifies the assumptions on the model of the overall distributed system and the possible attack scenarios.
Specifically, the P2P proxy forwards messages into the P2P network of the target system and listens for incoming messages. Similarly, the TOB proxy implements functions to input messages in a total-order broadcast protocol and read them once they have been ordered. Both proxies exploit gRPC API to define a client interface to insert messages into the network and a server interface to collect messages from the network.

\section{Evaluation}
\label{sec:evaluation}
We evaluate \NAME in relevant scenarios and compare the performance of the threshold schemes across multiple deployments. We benchmark \NAME under different workloads, in order to identify its capacity limits and possible bottlenecks.
\subsection{Setup}
The setup entails first an orchestrator node, which hosts a monitoring service, a client for invoking requests, and the benchmarking tools. Second is a network of nodes hosting the \NAME service, which we will refer to as the \emph{\NAMENETWORK} for the rest of the discussion. In more detail, the orchestrator node runs in a virtual machine with 16 GiB of RAM and 4 vCPU @ 2.2 GHz. Here, the benchmarking client, implemented in Rust, imports the schemes module (Sec.~\ref{sec:schemes}) of \NAME and implements the gRPC client-side \NAME API to create and schedule requests to the \NAMENETWORK according to the experiment parameters. The client also collects data and monitors the state of the network.
Furthermore, each node of the \NAMENETWORK runs on a dedicated virtual machine with 8 GiB of RAM and 2 vCPU @ 2.2 GHz. A \NAME instance runs encapsulated in a Docker container limited to 1 vCPU and 6 GiB of RAM of its dedicated VM in order to avoid interference with other services running on the same host: namely, a Prometheus monitoring server, which collects data about resource utilization, saturation conditions, and health status of the container.

\subsection{Parameters}
To measure the performance of \NAME in relevant real-world configurations,
we consider the following parameters: the geographical distribution of the nodes, the number of nodes composing the \NAMENETWORK, the request rate (load), and the request payload size.
We distinguish between a \emph{local} and a \emph{global} deployment. In the first, nodes reside in the same data center (FRA1), while in the second, they span different geographical regions (FRA1, SYD1, TOR1, SFO3).
Moreover, we choose three deployment sizes: a \emph{small} deployment with 7 nodes and a threshold of 3 (DO-7-L/G); a \emph{medium} deployment using 31 nodes, with a threshold of 11 (DO-11-L/G), and lastly, a \emph{large} deployment of 127 nodes, with a threshold of 43 (DO-127-L/G). 
The value of the threshold follows from the usual BFT assumption that fewer than one-third of nodes are faulty for $(t+1)$-out-of-$n$ schemes with $n=3t+1$.
Table~\ref{tab:deployments} presents a summary of the configurations with the measured average round-trip times between the nodes.
We also vary the request rate, starting from one req/s by repeatedly doubling it up to a highest rate that depends on the deployment, and we inject payloads of sizes that range from 256B to 4KiB.

\begin{table}[h!]
  \centering
  \renewcommand{\arraystretch}{1.2}
  \begin{tabular}{lcccc}
    \toprule
    Acronym    & Size   & Region(s)                         & \makecell{Network latency\\(ms)}& Max rate\\
    \midrule
    DO-7-L     & small  & FRA1                                & $\approx 0.65$       & 1024 req/s\\
    DO-7-G     & small  & \makecell{FRA1, SYD1,\\ TOR1, SFO3} & $\approx 100,43$     & 1024 req/s\\
    \midrule
    DO-31-L    & medium & FRA1                                & $\approx 0.65$       & 512 req/s\\
    DO-31-G    & medium & \makecell{FRA1, SYD1,\\ TOR1, SFO3} & $\approx 100,43$     & 512 req/s\\
    \midrule
    DO-127-G   & large  & \makecell{FRA1, SYD1,\\ TOR1, SFO3} & $\approx 100,43$     & 64 req/s\\
    DO-127-L   & large  & FRA1                                & $\approx 0.65$       & 64 req/s\\
    \bottomrule
  \end{tabular}
  \caption{Deployment configurations}
  \label{tab:deployments} 
\end{table}

\subsection{Metrics}
\label{sec:metrics}
\paragraph{Server-side latency} Latency is the time taken by the \NAMENETWORK to process a request, measured from when a server receives the request to when it generates the result for the threshold protocol. Consequently, it cannot be lower than the network latency between participants times the protocol latency i.e., the number of message transmission needed between the local processing steps.
Note that client-side latency would add one round-trip time for communication with the servers but is excluded.
In this analysis, we generally denote by $\mathcal{L}_k$ the value of the $k$-th percentile of the latency distribution across the requests processed by the whole \NAMENETWORK. When required, we specify the latency metric further: at the node level ($\mathcal{L}^{\text{node}}_{k}$) and the network level ($\mathcal{L}^{\text{net}}_{k}$).

Consequentially, we define the \emph{threshold latency} ($\mathcal{L}^{\text{net}}_{\theta}$) as the $\theta$-th percentile of the latency distribution across nodes. Here, $\theta$ is calculated as $\theta = \frac{t+1}{n}\cdot 100$, where $t$ is the threshold parameter of the scheme under test and $n$ is the size of the network, i.e., roughly $\theta=34$. This metric indicates how quickly the \NAMENETWORK can produce the final result. Notably, the gap between $\mathcal{L}^{\text{net}}_{95}$ and $\mathcal{L}^{\text{net}}_{\theta}$ highlights residual delays caused by slower nodes, which continue to affect overall performance even after the primary computation is completed.
We select this metric because considering the average alone may leave out important effects that are relevant in practical deployments. 
Thus, we define a \emph{Residual Delay Factor} as
\begin{equation*}
  \delta_{\text{res}} = \frac{\mathcal{L}^{\text{net}}_{95} - \mathcal{L}^{\text{net}}_{\theta}}{\mathcal{L}^{\text{net}}_{\theta}},
\end{equation*}
which quantifies the extent to which slow nodes influence $\mathcal{L}^{\text{net}}_{95}$. 
Given that $\mathcal{L}^{\text{net}}_{95} \geq \mathcal{L}^{\text{net}}_{\theta}$, the value $\delta_{\text{res}}$ is in $\mathbb{R}^{\geq 0}$, therefore, the following interpretation: for values of $\delta_{\text{res}}$ approaching $1$, the residual delay introduced by slow nodes has a small impact on the system's performance. In this range, the overhead remains within twice $\mathcal{L}^{\text{net}}_{\theta}$, representing a tolerable shift equivalent to a single processing unit in the protocol's operation. Conversely, as $\delta_{\text{res}}$ gets much larger, the impact of slow nodes becomes significant, creating considerable delays and potentially degrading system responsiveness in approaching full load.
Further, we introduce the \emph{Latency Fairness Index} as
\begin{equation*}
  \eta_{\theta} = \frac{\mathcal{L}^{\text{net}}_{\theta}}{\mathcal{L}^{\text{net}}_{95}},
\end{equation*}
which evaluates how evenly the load is distributed among nodes based on their contribution to latency. The index reflects how closely the performance of the $(t+1)$-quorum of fastest nodes lies relative to $95\%$ of the nodes; it ranges from $0$ to $1$. Values of $\eta_{\theta} \in [0.5, 1]$ indicate a fair balance across the nodes, values close to $1$ suggest that most nodes perform similarly, whereas values around $0.5$ suggest an imbalance within a factor two. Values much lower than $0.5$ indicate significant imbalance, where a subset of fast nodes dominates the protocol and leaves many slower nodes lagging behind. This effect may lead to bottlenecks, reduced throughput, or even vulnerabilities in protocol operation.
It is worth noting that the two metrics, $\delta_{\text{res}}$ and $\eta_{\theta}$, are inversely related. As $\delta_{\text{res}}$ increases, indicating a greater impact of slow nodes on system performance, $\eta_{\theta}$ decreases, reflecting a worsening balance between the nodes' contribution to the protocol. Conversely, a small $\delta_{\text{res}}$ suggests minimal impact of the slow nodes, and the fairness index $\eta_{\theta}$ grows accordingly, indicating a more balanced contribution from all nodes in the network. 

\paragraph{Throughput}  Throughput denotes the number of requests processed by the system per time unit and measures the system's capacity. We estimate the throughput as the ratio between the number of requests processed by the \NAMENETWORK and the elapsed time between the first and last correctly processed request. If processing extends beyond the experiment's duration, a grace period of up to 10\% is considered. Conversely, when the load is high, and requests remain unprocessed, the total experiment duration is used as the time unit to ensure consistency in the metric.

\begin{table}[h!]
  \centering
  \renewcommand{\arraystretch}{1.2}
  \begin{tabular}{llcc}
    \toprule
    Scheme & \makecell{Arithmetic \\Structure} & \makecell{Key \\lenght (bit)} & \makecell{Communication \\complexity} \\
    \midrule
    SG02  & EC (Ed25519)       & 256    & $\CO(n)$   \\
    BZ03  & EC (Bn254)         & 254    & $\CO(n)$   \\
    SH00  & RSA                & 2048   & $\CO(n)$   \\
    BLS04 & EC (Bn254)         & 254    & $\CO(n)$   \\
    KG20  & EC (Ed25519)       & 256    & $\CO(n^2)$ \\
    CKS05 & EC (Ed25519)       & 256    & $\CO(n)$   \\
    \bottomrule
  \end{tabular}
  \caption{Schemes' parameters benchmark setup} 
  \label{tab:schemessetup}
\end{table}

\subsection{Methodology}
\label{sec:methodology}
The \NAMENETWORK is restarted before each experiment to ensure a clean state. The tested schemes and configurations are summarized in Table~\ref{tab:schemessetup}. For each scheme relying on elliptic curve cryptography, we indicate the curve used for the experiments. We additionally list the key lenght and the asymptotic communication complexity of the scheme. Notably, only KG20 requires two rounds of communication. Though KG20 supports batch precomputation in a first round, we consider the worst-case scenario and measure it across both rounds.

We assume a setup phase during which a trusted dealer distributes the key material for all schemes. Every threshold protocol implemented in \NAME performs both a share verification, to ensure the correctness of the shares, and a result verification upon assemble, to ensure the correctness of the final output. %
This ensures a fair comparison of the schemes in the following analysis.

Firstly, we conduct a \emph{capacity test} of the system under the different deployments by increasing the request rate (in factors of two) until the system reaches saturation, starts to degrade in latency, or exhibits failures.  
Table~\ref{tab:deployments} shows the maximum request rate chosen for each deployment.
Every experiment lasts one minute.
The goal is to identify the maximum sustainable throughput, or \emph{usable capacity}, the system can handle without significant performance degradation. The \emph{knee capacity} is estimated as the request rate at which the ratio of throughput to latency is maximized, marking the optimal efficiency point. The range between the knee and usable capacity defines the system's operating region, where it performs reliably under load.
Here, we consider latency as the $\mathcal{L}_{95}$ value across all requests processed.
Secondly, we choose the medium-scale global deployment (DO-31-G) to conduct longer experiments of five minutes. We aim to observe the system in a steady state and check if the corrisponding workload can be handled over a longer period.
Here, we evaluate the \NAMENETWORK performance closely, focusing on $\mathcal{L}^{\text{net}}_{50}$, $\mathcal{L}^{\text{net}}_{\theta}$, and $\mathcal{L}^{\text{net}}_{95}$, and the derived metrics $(\delta_{\text{res}},\eta_\theta)$ computed from $\mathcal{L}^{\text{node}}_{95}$. 
Finally, we repeat the 5-minutes-long experiment varying the size of the request payload, from 256B to 4KiB, to evaluate the impact of the size of the data on the system's latency.

\begin{figure*}[t!]
  \centering
  \begin{subfigure}[b]{0.49\textwidth}
      \centering
      \includegraphics[width=\columnwidth]{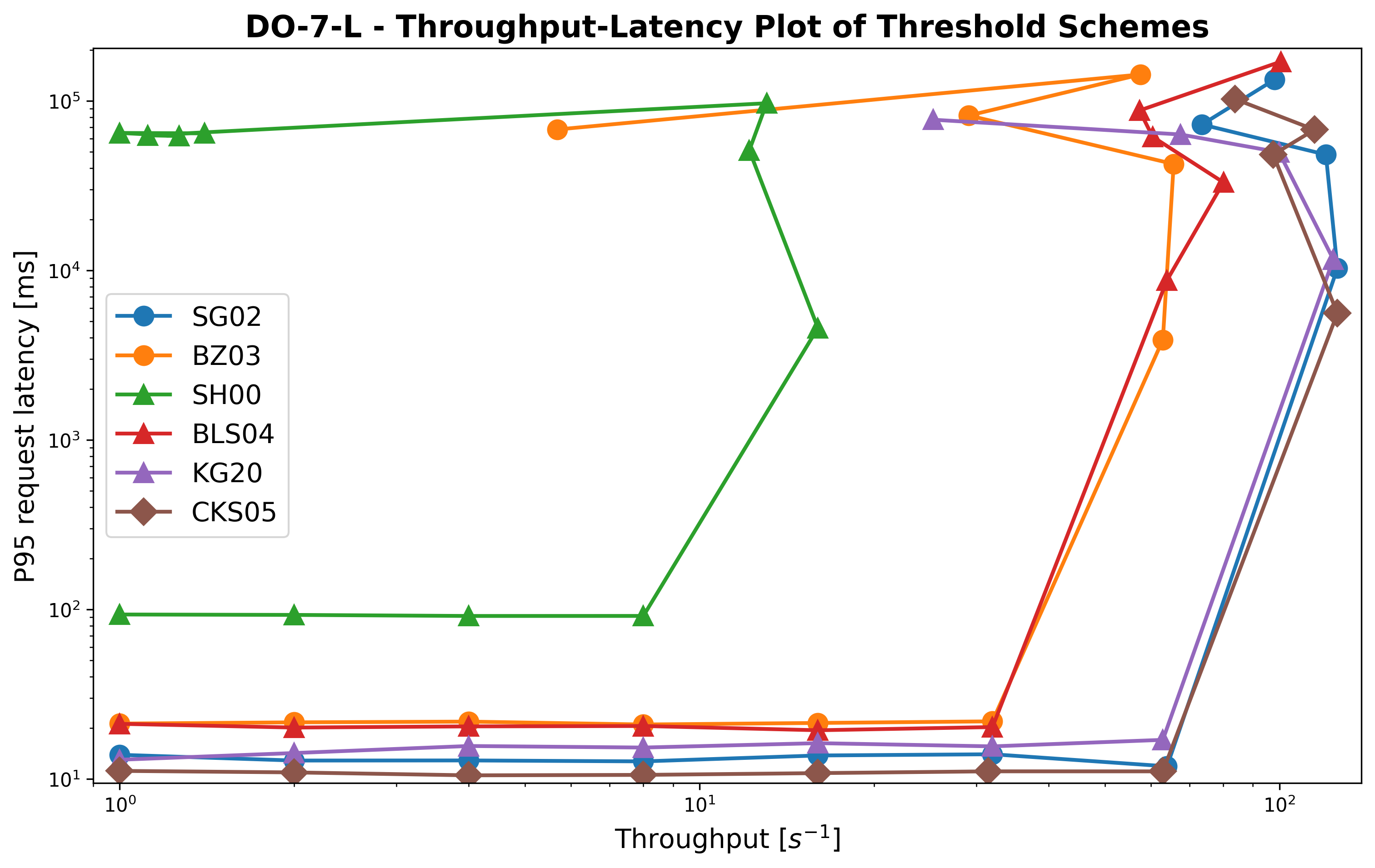} %
      \label{fig:7_regional}
  \end{subfigure}
  \centering
  \begin{subfigure}[b]{0.49\textwidth}
      \centering
      \includegraphics[width=\columnwidth]{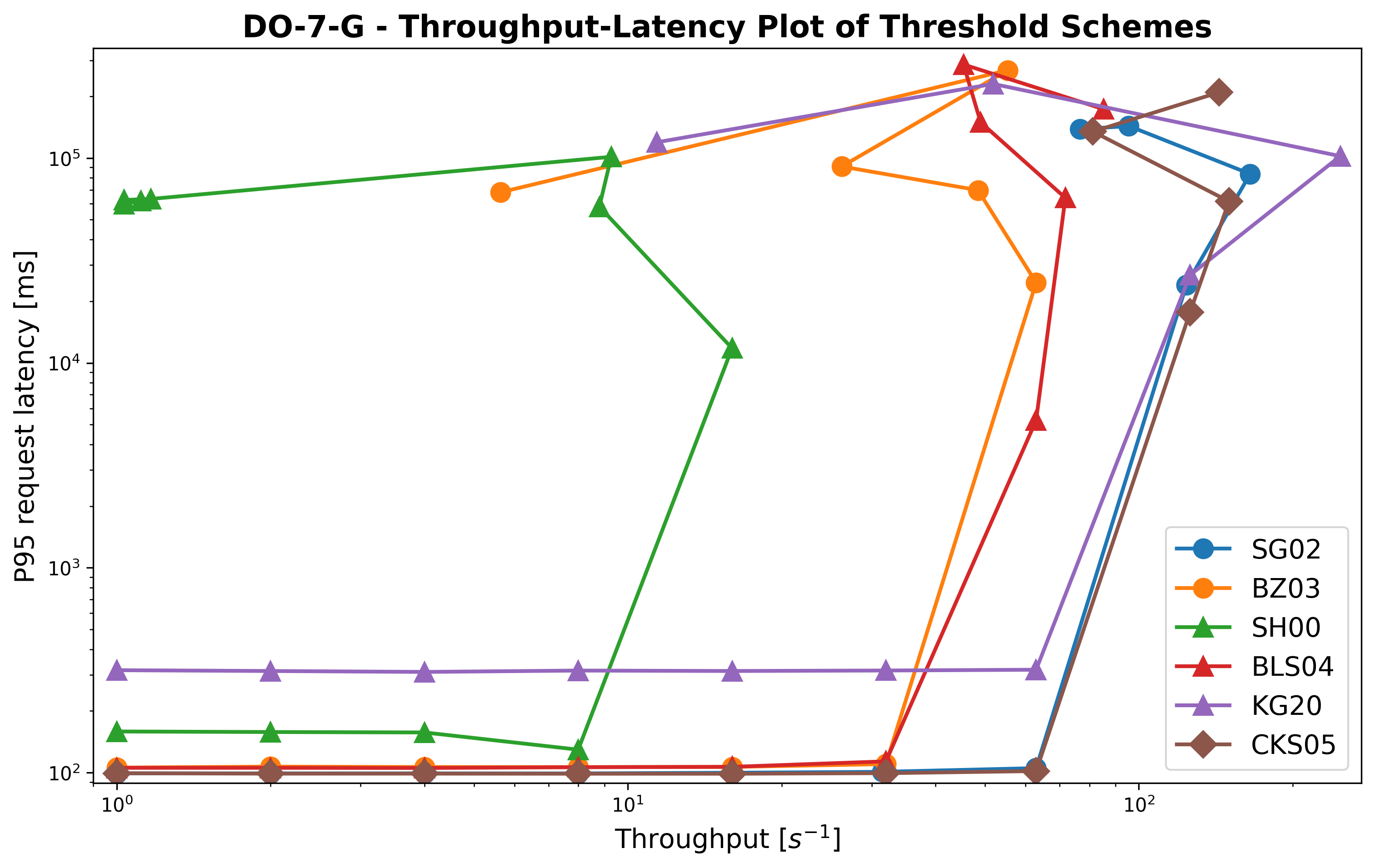} %
      \label{fig:7_global}
  \end{subfigure} 
  \centering
  \begin{subfigure}[b]{0.49\textwidth}
      \centering
      \includegraphics[width=\columnwidth]{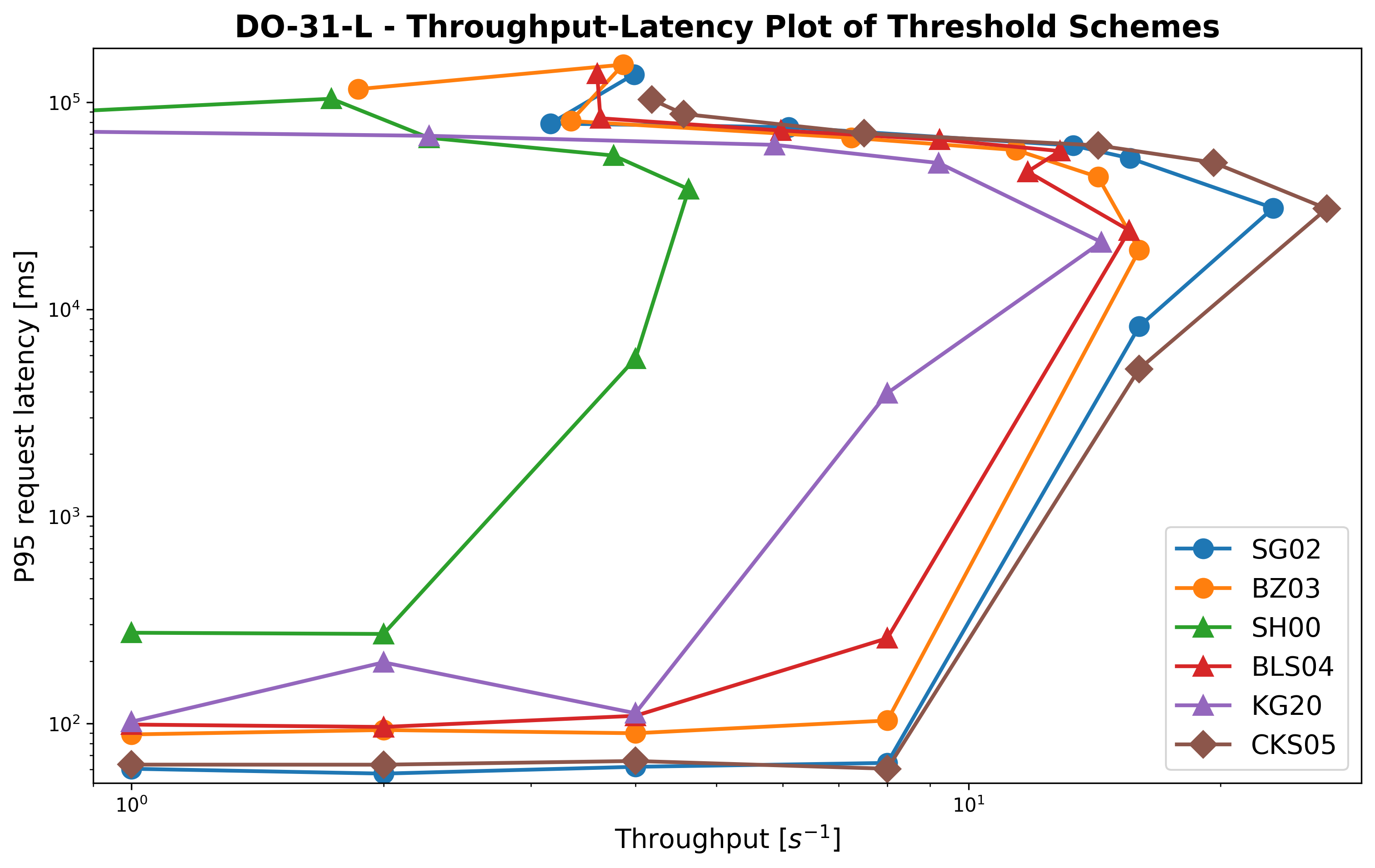} %
      \label{fig:31_regional}
  \end{subfigure}
  \centering
  \begin{subfigure}[b]{0.49\textwidth}
    \centering
  \includegraphics[width=\columnwidth]{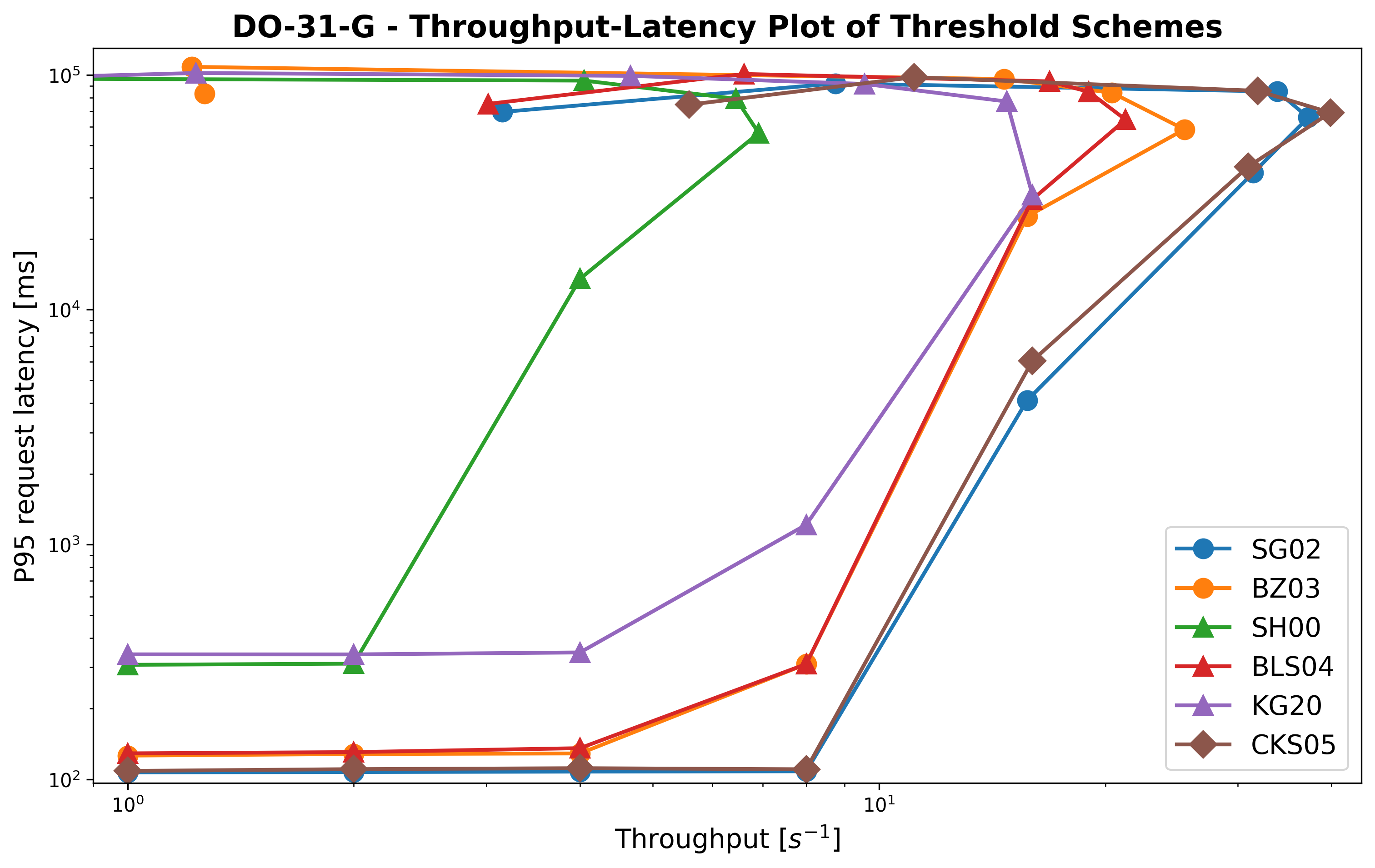} %
  \label{fig:31_global}
  \end{subfigure}
  \centering
  \begin{subfigure}[b]{0.49\textwidth}
    \centering
    \includegraphics[width=\columnwidth]{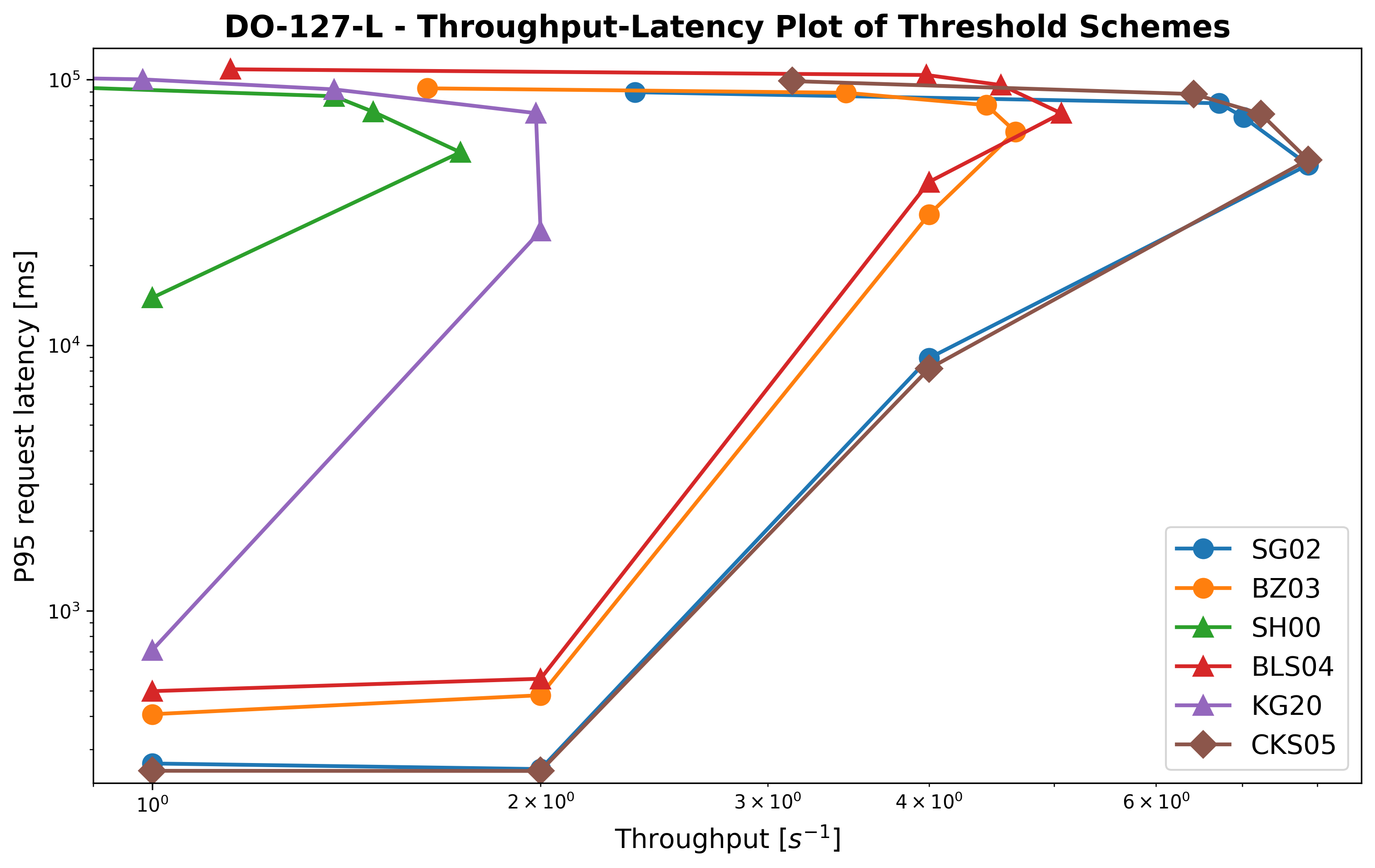} %
    \label{fig:127_regional}
  \end{subfigure}
  \centering
  \begin{subfigure}[b]{0.49\textwidth}
    \centering
    \includegraphics[width=\columnwidth]{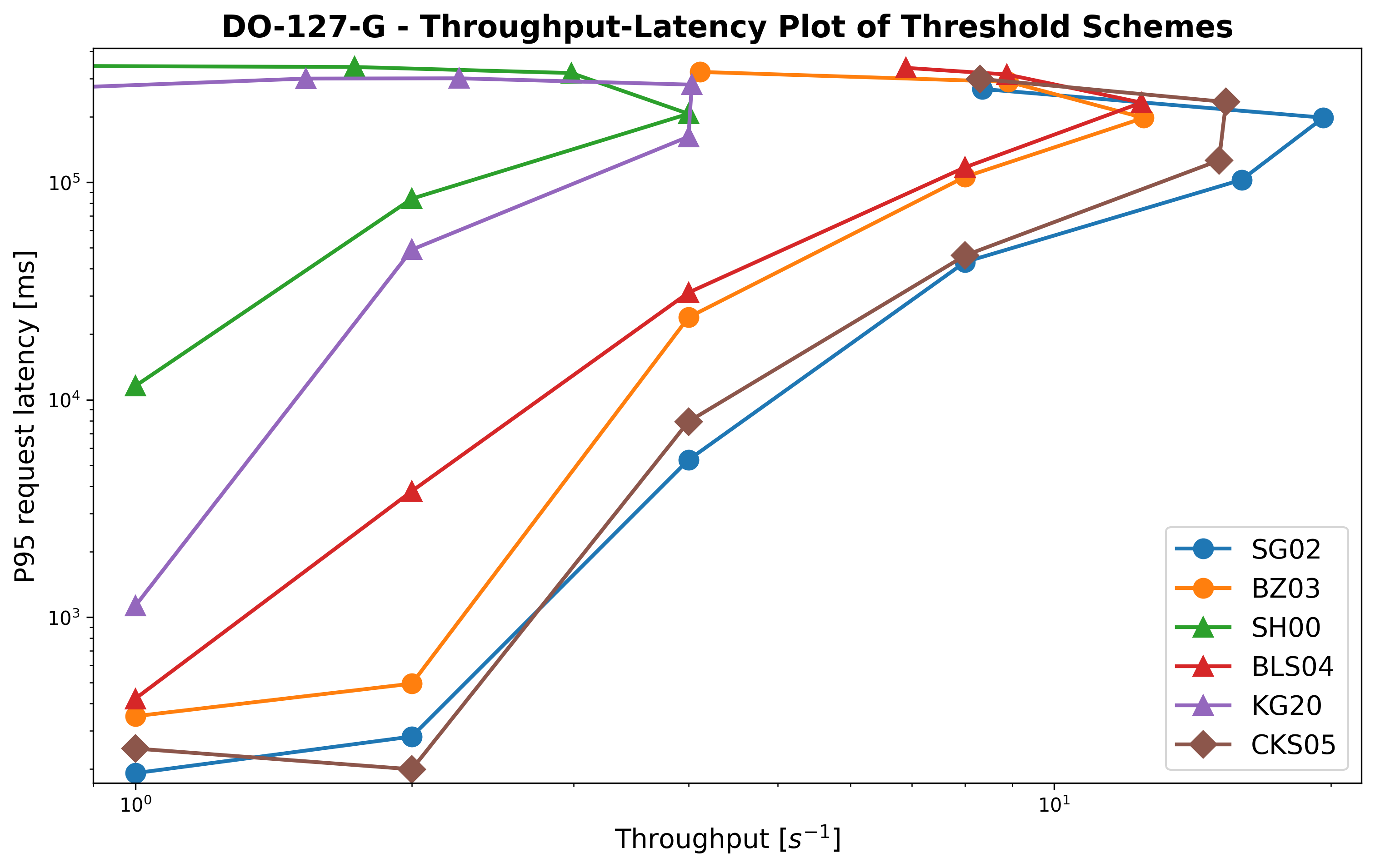} %
    \label{fig:127_global}
  \end{subfigure}
  \caption{Server-side Throughput-Latency graphs}
  \label{fig:server_latencies}
\end{figure*}

\subsection{Results}
We discuss the results of our empirical evaluation following the described methodology. 
\paragraph{Capacity test} Figure~\ref{fig:server_latencies} presents the capacity test results as throughput-latency graphs, organized by deployment size (rows) and geographical configuration (columns) for each size.
For each data series corresponding to a certain load and scheme, we consider $\mathcal{L}_{95}$ as function of the throughput estimated as described in~Section~\ref{sec:metrics}.
Observing the graph, we can quickly identify the \emph{knee point} as the last point before latency increases non-linearly due to resource contention (CPU, network bandwidth, I/O). The knee point is the graphical equivalent of the knee capacity. The rightmost value of the graph, in contrast, indicates the maximum throughput that the system can handle before degrading. 
The latency values range from a lower bound imposed by network latency to an upper bound due to the experiment time ($\approx 60s$), as we calculate latency solely for completed requests.

he measurements show that the category of a scheme (i.e., cipher, signature, randomness) appears to be less relevant than the underlying cryptographic assumption. Therefore, we compare schemes based on the assumptions they rely on: SG02, KG20 and CKS05 are based on the Diffie-Hellman assumption over elliptic curves (ECDH), BLS04 and BZ03 rely on pairings, finally SH00 is based on RSA assumption (recall Tab.~\ref{tab:schemes}). Following this grouping, we expect ECDH to be less computationally intensive than pairings, and RSA to be the most computationally intensive.
The second dimension to consider is communication complexity. 

Considering these distinctions, we delve into the results of the various deployments.
We first look at the \emph{geographical distribution} parameter, comparing the local (left comumn) with the global (right column) deployments.
As expected, we observe this parameter significantly impacts latency, due to the higher network latency introduced by the global deployment. The knee point of the schemes remains generally unchanged when moving from the local to the global setup across all deployment sizes; the only exception is BLS04 going from a knee point of 2 req/s (DO-127-L) to 1 req/s (DO-127-G). This indicates that the achievable capacity depends primarily on the local computation power; in other words, throughput remains unaffected by geographical distribution.

The \emph{deployment size}, i.e., the number of nodes participating in the protocol, instead plays a crucial role.  
In the first row of Fig.~\ref{fig:server_latencies}, deployments DO-7-L and DO-7-G show pronounced differences among the schemes, emphasizing the impact of local cryptographic computations. In particular, CKS05, KG20, and SG02, relying on ECDH, perform better than the BLS04 and BZ03 schemes, which rely on pairings. Notably, KG20, despite requiring two rounds of communication, achieves faster performance in these settings than SH00, which involves computationally heavy RSA signature operations. Thus, for small deployments (7 nodes), local cryptographic computation have a greater impact than communication overhead introduced by an additional round. Regarding knee point values, the DH-based schemes reach knee point at 64 req/s, pairing-based at 32 req/s, and the RSA-based one at 8 req/s.
In DO-31-L and DO-31-G (middle row), performance differences between the schemes become less pronounced due to the increasing number of participants, which directly affects the threshold parameter and, consequently, the overall latency. For all the non-interactive schemes except SH00, the knee point is reduced by a factor of $2^3$ e.g., decreasing from 64 req/s to 16 req/s for SG02. SH00 here reaches knee point at 4 req/s. 
As opposed to the previous case, KG20 reaches knee point already at 8 req/s, revealing the impact of the second communication round as the deployment size scales. 
In DO-127-L and DO-127-G (bottom row), differences are nearly negligible, as system performance becomes predominantly determined by the number of participants and the network latency among them. The knee point drops further to 2 req/s for all schemes, except for SH00 and KG20, which reaches it at 1 req/s.

In summary, the capacity test reveals that system performance depends on cryptographic assumptions and participant number. The former dominates in small deployments, while the latter becomes crucial in larger ones.
The relative order of the non-interactive schemes remains consistent: ECDH-based schemes outperform pairing-based schemes, and RSA-based schemes are consistently the slowest. KG20 benefits from its mathematical assumption (ECDH) in small deployments but shows the impact of its second communication round as deployment size increases.

\begin{figure*}[t!]
  \begin{subfigure}[b]{0.49\textwidth}
  \centering
    \centering
    \includegraphics[width=\columnwidth]{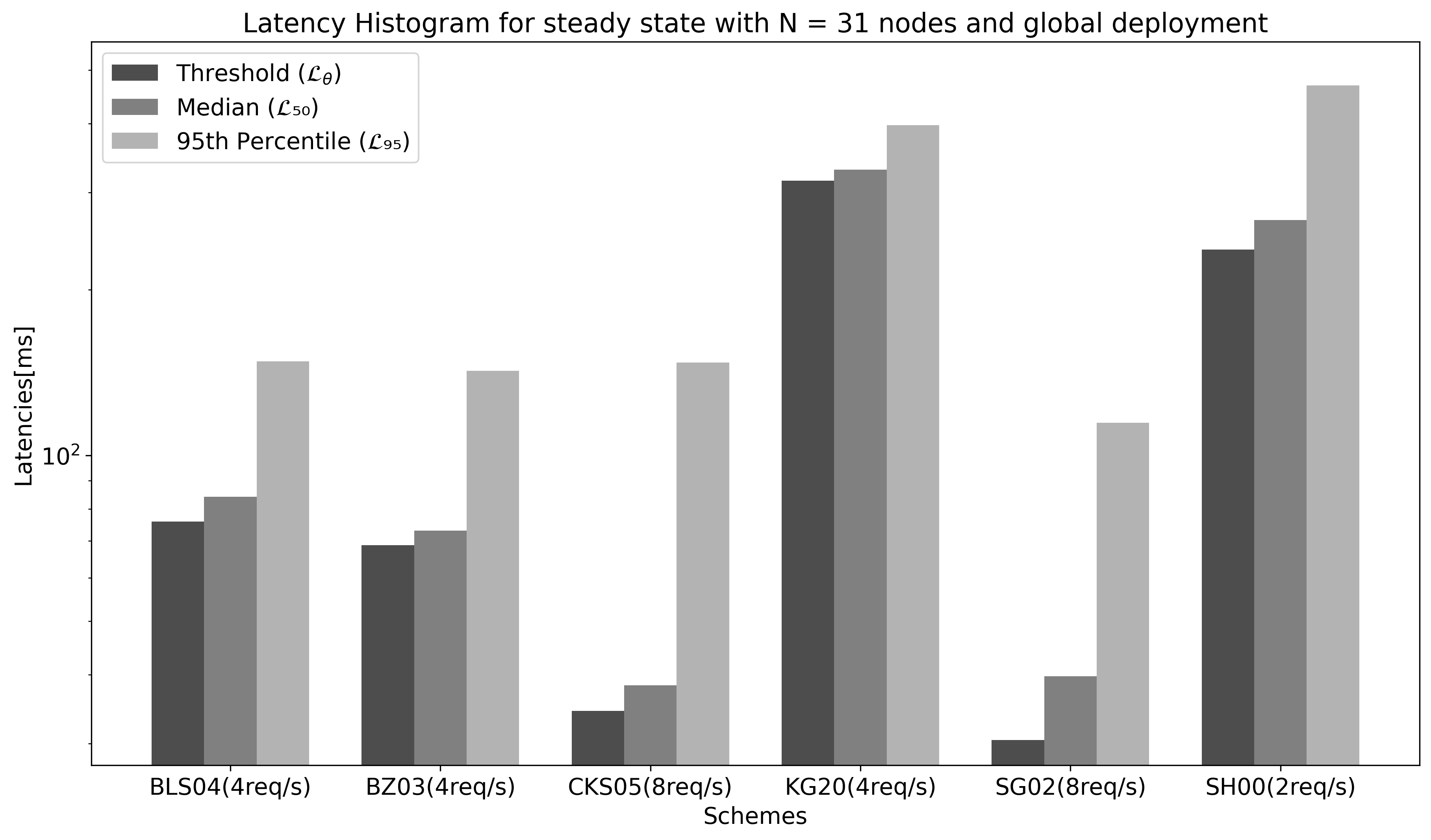}
    \caption{Percentile comparison of latency ($\mathcal{L}_{\theta}$, $\mathcal{L}_{50}$,  $\mathcal{L}_{95}$)}
    \label{fig:histogram_comparison} %
  \end{subfigure}
  \begin{subfigure}[b]{0.49\textwidth}
    \centering
    \includegraphics[width=\columnwidth]{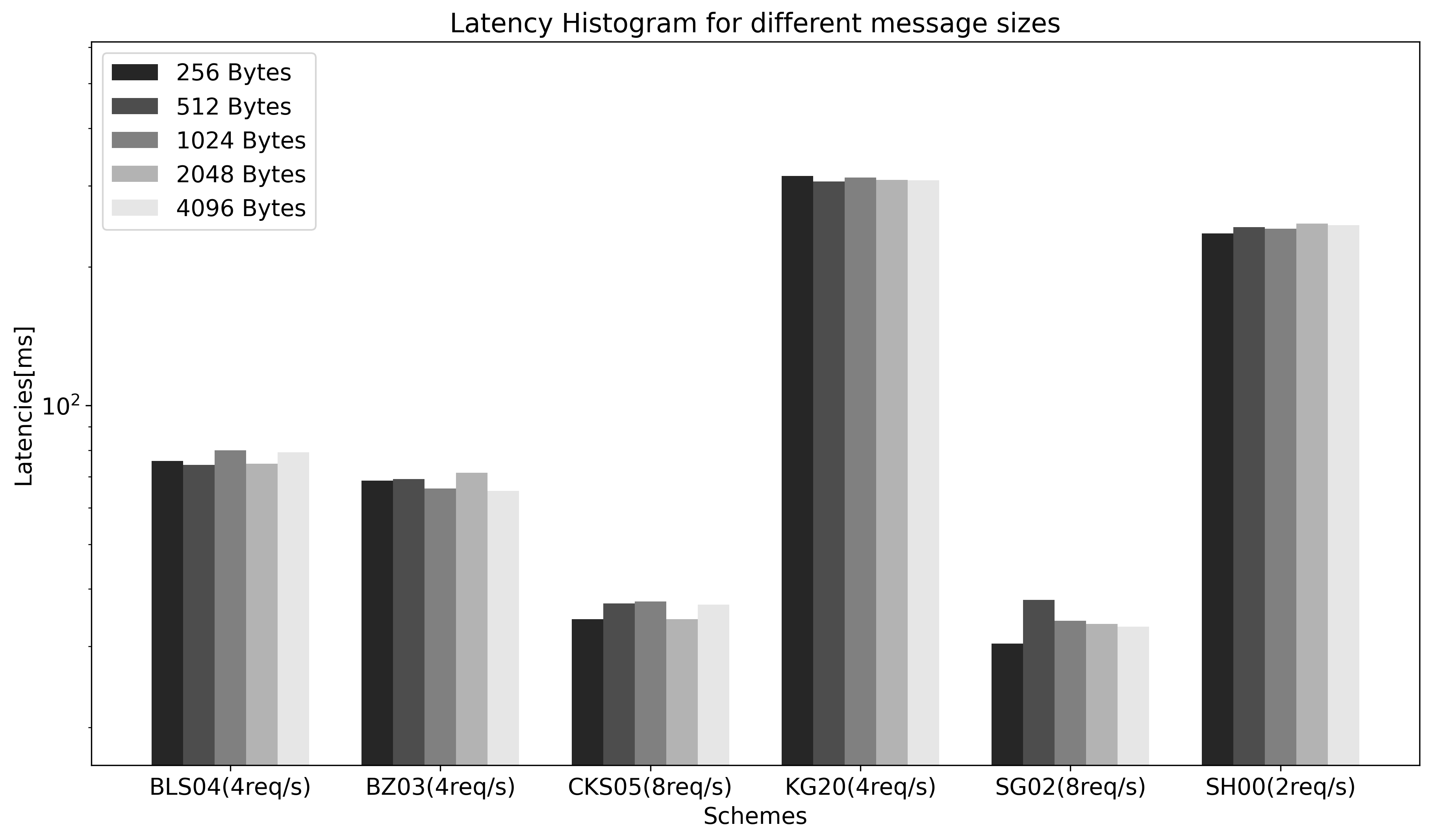}
    \caption{$\mathcal{L}_{\theta}$ per message size}
    \label{fig:message_size_comparison}
  \end{subfigure}
  \label{fig:5-minutes-long}
  \caption{Five-minute experiments at knee capacity.}
\end{figure*}

\paragraph{Steady-state analysis}
We conduct next a five-minutes-long experiment in a steady state using the DO-31-G deployment, studing the latency distribution among the nodes (Fig.~\ref{fig:histogram_comparison}) and the impact of the request payload size (Fig.~\ref{fig:message_size_comparison}).
We run this experiment under a load that corresponds to the knee capacity identified earlier and consider this the steady state.
Table~\ref{tab:schemesperformance} summarizes the computed knee capacities for each scheme in DO-31-G, together with the newly introduced latency metrics.
We are interested in evaluating latency differences across nodes, and the impact of the threshold parameter on the system's performance.
We calculate the latency metrics as described in Section~\ref{sec:methodology}.
Here, we use $\mathcal{L}^{\text{net}}_{\theta}$ to estimate the overall request processing time, specifically when $t+1$-out-of-$n$ nodes have completed the computation. As shown in Figure~\ref{fig:histogram_comparison}, schemes where the local computations are more expensive, such as SH00 and KG20, exhibit higher values. 
The clearly visible gap increase from $\mathcal{L}^{\text{net}}_{\theta}$ to $\mathcal{L}^{\text{net}}_{95}$, on the other hand, reflects how long slow nodes continue to impact the \NAMENETWORK with residual messages from an already processed request.
The numerical results in terms of $\delta_\text{res}$ and $\eta_{\theta}$ are depicted in Table~\ref{tab:schemesperformance}. 

\begin{table}[tbh]
    \centering
    \renewcommand{\arraystretch}{1.2}
    \begin{tabular}{lccc}
      \toprule
      Scheme & \makecell{Knee Capacity} & \makecell{$\delta_\text{res}$} & \makecell{$\eta_{\theta}$}\\
      \midrule
      SG02  & 8 req/s & 2.764 & 0.266\\
      BZ03  & 4 req/s & 1.074 & 0.482\\
      SH00  & 2 req/s & 0.986 & 0.503\\
      BLS04 & 4 req/s & 0.953 & 0.512\\
      KG20  & 4 req/s & 0.260 & 0.793\\
      CKS05 & 8 req/s & 3.285 & 0.233\\
      \bottomrule
    \end{tabular}
    \caption{Performance summary, using DO-31-G} 
    \label{tab:schemesperformance}
  \end{table}

This gap is highest for the non-interactive DH-based schemes and lower for pairing-based ones.
If, on one side, the small $\CL^{\text{net}}_\theta$ reflect fast completion of the computation, the high $\delta_{\text{res}}$ values indicate that the network is affected by the slow nodes.
The RSA-based schemes exhibit roughly the same gap as pairing-based schemes, indicating a reasonable balance between the local computation and the communication overhead. This is also reflected in their $\eta_{\theta}$ values around $0.5$.
KG20, on the other hand, presents the smallest gap ($\delta_{\text{res}} \approx 0.26$). We suspect the reason is the second round of communication as well as the fixed signing group requirement,  i.e., the protocol will wait for the contributions of all nodes in the a-priori defined group to complete. The index $\eta_{\theta}$ indicates that the protocol is not skewed by the threshold, which we expect from the protocol design.

Finally, Figure~\ref{fig:message_size_comparison} shows the impact of the request payload size on the $\mathcal{L}^{\text{res}}_{\theta}$ value. As expected, the payload size does not significantly affect latency throughout the schemes. This is because the signature and the randomness schemes hash the message first, and because the cipher uses hybrid encryption.

\section{Related work}
\label{sec:relwork}
In this section, we mention some directly related recent work on practical integration of threshold cryptography, excluding theoretical contributions as the literature on threshold cryptography is vast.
Two orthogonal lines of research are relevant: integration of threshold cryptography into distributed systems to enhance security and relax system-wide trust assumptions, and easier implementation of more complex multi-round protocols, relying on the coordination of replicated systems. 

As for the first, 
F3B~\cite{DBLP:conf/aft/ZhangMQBEF23} is an extension to a blockchain that protects against front-running attacks through threshold encryption.  It employs two committees: one for consensus and one for decryption, and the two do not necessarily overlap.  Since the solution is tailored to Ethereum, it is specific to its consensus model, exploiting long finality times. 
Shutter Network~\cite{shutter} provides an identity-based threshold decryption service for Ethereum smart contracts. Leveraging an ad-hoc committee, the network allows in every epoch decryption of encrypted transactions on the blockchain in the previous epoch, resulting in a delay of one epoch. Shutter relies on a Tendermint-based atomic broadcast channel for communication. This ensures fault-tolerance when the threshold parameter is chosen accordingly, sacrificing latency due to ordering.
The Internet Computer~\cite{DBLP:conf/podc/CamenischDHPS022} demonstrates the successful deployment of threshold protocols in practice, using secure randomness for leader selection and quorum certificates for consensus. However, the cryptographic methods are integrated closely with the whole Internet Computer platform.
Also Aptos recently released a design for blockchain-native threshold cryptography~\cite{DBLP:journals/corr/abs-2407-12172}.  It provides cryptographic randomness for the platform, but it is also intertwined with the consensus mechanism to optimize performance.

A separate line of work has proposed to use blockchains as a coordination layer for implementing complex, multi-round protocols. For instance, Ferveo~\cite{DBLP:journals/iacr/BebelO22} introduces a DKG protocol that relies on BFT-based blockchain synchronization, enabling threshold encryption for mempool privacy. Similarly, CHURP~\cite{DBLP:conf/ccs/MaramZWLZJS19} focuses on secure reconfiguration and resharing strategies, while Scrape~\cite{DBLP:conf/acns/CascudoD17} proposes randomness beacon generation using blockchain as a communication medium.

Of a different interest is the performance of threshold schemes in practical systems, with measurements that go beyond microbenchmarks. A recent benchmarking efforts compares multiple libraries with respect to their effectiveness for BFT replication protocols~\cite{vonseck2024thresh}. 
Similar to our evaluation, this study demonstrates the crucial importance of the environment and network deployment on the observed performance of threshold cryptography.
 
\section{Conclusion}  
\label{sec:conclusion}  

\NAME aligns with the goals of the ongoing NIST standardization~\cite{nist-tc} offering a practical generalized framework for threshold cryptography. It overcomes the limitations of crypto libraries by providing a service with a unified interface that abstracts detail of specific schemes, and it can readily be integrated in existing distributed platforms. 

The dual API interface allows for both fine-grained access to the cryptographic core and high-level access to the threshold protocols in a black-box manner.
The modular design of the Threshold Round Interface allows the management code to be completely independent of specific protocol details favoring service extensibility to support new, potentially multi-round protocols. 
Similarly, the network interface can be implemented on different platforms, depending on the hosting environment, and plugged in the service without affecting the core functionality.

The modularity and software design choices of \NAME grant also a perfect environment to develop, test and compare different threshold schemes under the same conditions, guiding the choice of the most suitable scheme for a specific application or deployment.
Our evaluation provides insights into the performance of threshold cryptography in a practical system, showing which parameters are relevant to study and which metrics are meaningful to system-wide performance, enhancing and complementing the limited results of microbenchmarks.
The open source implementation of \NAME is available at \url{https://github.com/cryptobern/thetacrypt}. 

\bibliographystyle{ieeesort}
\bibliography{dblpbibtex, references}

\end{document}